\documentclass[a4paper,12pt]{article}

\usepackage{indentfirst}

\usepackage{amsfonts,amsmath,amssymb,bm,a4wide,graphicx,cite}

\fussy

\begin{document}

  \date{}
  \title{Generalized Dirac-Pauli Equation and Spin Light
  of Neutrino in Magnetized Matter}

\author{ Alexander Grigoriev \footnote{e-mail: ax.grigoriev@mail.ru},
Alexander Studenikin \footnote{e-mail: studenik@srd.sinp.msu.ru},
Alexei Ternov \footnote{e-mail:A\_Ternov@mail.ru}
\\
\small{\it Physics Faculty of Moscow State University, 119992
Moscow, Russia}}

\date{}
\maketitle

\begin{abstract}We consider propagation of a massive
neutrino in matter within the quantum approach based on the two
equations for the neutrino field: the first one is the Dirac-Pauli
equation for a massive neutrino in an external magnetic field
generalized on the inclusion of effects of the background matter;
the second one is the modified Dirac equation derived directly
from the neutrino-matter interaction Lagrangian. On the basis of
these two equations the quantum theory of a neutrino moving in the
background matter is developed (the exact solutions of these
equations are found and classified over the neutrino spin states,
the corresponding energy spectra are also derived). Using these
solutions we study within the quantum approach the spin light of
neutrino ($SL\nu$) in matter with the effect of a longitudinal
magnetic field being also incorporated. In particular, the $SL\nu$
radiation rate and total power are derived. The use of the
generalized Dirac-Pauli equation also enables us to consider the
$SL\nu$ in matter polarized under the influence of strong magnetic
field.
\end{abstract}

\section{Introduction}
Recently in a series of our papers
\cite{EgoLobStuPLB00LobStuPLB01,DvoStuJHEP02,StuPAN04} we have
developed the quasi-classical approach to the massive neutrino
spin evolution in the presence of external electromagnetic fields
and background matter. In particular, we have shown that the well
known Bargmann-Michel-Telegdi (BMT) equation \cite{BarMicTelPRL59}
of the electrodynamics can be generalized for the case of a
neutrino moving in the background matter and being under the
influence of external electromagnetic fields. The proposed new
equation for a neutrino, which simultaneously accounts  for the
electromagnetic interaction with external fields and also for the
weak interaction with particles of the background matter, was
obtained from the BMT equation by the following substitution of
the electromagnetic field tensor $F_{\mu\nu}=(\bf E,\bf B)$:
\begin{equation}\label{sub}
F_{\mu\nu} \rightarrow E_{\mu\nu}= F_{\mu\nu}+G_{\mu\nu},
\end{equation}
where the tensor $G_{\mu\nu}=(-{\bf P},{\bf M})$ accounts for the
neutrino interactions with particles of the environment. The
substitution (\ref{sub}) implies that in the presence of matter
the magnetic $\bf B$ and electric $\bf E$ fields are shifted by
the vectors $\bf M$ and $\bf P$, respectively:
\begin{equation}
\bf B \rightarrow \bf B +\bf M, \ \ \bf E \rightarrow \bf E - \bf
P. \label{11}
\end{equation}
We have also shown  how to construct the tensor $G_{\mu \nu}$ with
the use of the neutrino speed, matter speed, and matter
polarization four-vectors.

Within the developed quasi-classical approach to the neutrino spin
evolution we have also considered
\cite{LobStuPLB03,DvoGriStuIJMP04,LobStuPLB04} a new type of
electromagnetic radiation by a neutrino moving in the background
matter in the presence of electromagnetic and/or gravitational
fields which we have named the "spin light of neutrino" ($
SL\nu$). The $ SL\nu$ originates, however, from the quantum spin
flip transitions and for sure it is important to revise the
calculations of the rate and total power of the $ SL\nu$ in matter
using the quantum theory. Note that within the quantum theory the
radiation emitted by a neutrino moving in a magnetic field was
also considered  in \cite{BorZhuTer88}.

In this paper we should like to present a reasonable step forward,
which we have made recently \cite{StuTer0410296_97}, in the study
of the neutrino interaction in the background matter and external
fields. The developed quantum theory of a neutrino motion in the
presence of the background matter is based on the two equations
for the neutrino wave function. The first equation is obtained in
the generalization of the Dirac-Pauli equation of the quantum
electrodynamics under the assumption that matter effects can be
introduced through the substitution (\ref{sub}). The second of
these equations is derived directly from the neutrino interaction
Lagrangian averaged over the particles of the background.  In the
limit of the constant matter density, we get the exact solutions
of these equations, classify them over the neutrino helicity
states and determine the energy spectra, that depend on the
helicity.

Although the neutrino energy spectra in matter correspondent to
the two equations are not equal, the difference of energies of the
opposite neutrino helicity states, predicted in the linear
approximation in the matter density by the two equations, are
equal. Then with the use of the obtained neutrino quantum states
in matter we develop the quantum theory of the $ SL\nu$ and
calculate the emitted photon energy, the rate and power of the
radiation in matter accounting for the emitted photons
polarization. The generalized Dirac-Pauli equation enables us to
include the contribution from the longitudinal magnetic field and
also to account for the effect of the matter polarization.

\section {Neutrino wave function and energy spectrum in matter}

In this section we discuss the two quantum equations for a massive
neutrino wave function in the background matter. The two wave
functions and energy spectra, that correspond to these two
equations, are not the same. However, the results for the $SL\nu$
photon energy, the rate and the power obtained on the basis of
these two equations in the lowest approximation over the matter
density (see Section 3), are equal.
\subsection{Modified Dirac-Pauli equation for neutrino in matter}

To derive a quantum equation for the neutrino wave function in the
background matter we start with the well-known Dirac-Pauli
equation for a neutral fermion with non-zero magnetic moment. For
a massive neutrino moving in an electromagnetic field $F_{\mu
\nu}$ this equation is given by
\begin{equation}\label{D_P}
\Big( i\gamma^{\mu}\partial_{\mu} - m -\frac{\mu}{2}\sigma ^{\mu
\nu}F_{\mu \nu}\Big)\Psi(x)=0,
\end{equation}
where $m$ and $\mu$ are the neutrino mass and magnetic
moment\footnote{For the recent studies of a massive neutrino
electromagnetic properties, including discussion on the neutrino
magnetic moment, see Ref.\cite{DvoStuPRD_JETP04}}, $\sigma^{\mu
\nu}=i/2 \big(\gamma^{\mu }\gamma^{\nu}-\gamma^{\nu}
\gamma^{\mu}\big)$.

Now let us consider the case of a neutrino moving in the presence
of matter without any electromagnetic field in the background. Our
goal is to study the spin-light photon emission in the process of
the neutrino transition between the two quantum states with
opposite helicities in the presence of matter. Since the
calculation of the $SL\nu$ rate and power are performed below (in
Section 3) within the lowest approximation over the density of the
background matter, we are interested now in the difference of
energies of the two neutrino states with opposite helicities in
the presence of matter. The quantum equation for the neutrino wave
function, appropriate for this task, can be obtained from
(\ref{D_P}) with application of the substitution (\ref{sub}) which
now becomes
\begin{equation}\label{sub_1}
  F_{\mu \nu}\rightarrow G_{\mu \nu}.
\end{equation}
Thus, we get the quantum equation for the neutrino wave function
in the presence of the background matter in the form
\cite{StuTer0410296_97}
\begin{equation}\label{D_P_matter}
\Big( i\gamma^{\mu}\partial_{\mu} - m -\frac{\mu}{2}\sigma ^{\mu
\nu}G_{\mu \nu}\Big)\Psi(x)=0,
\end{equation}
that can be regarded as the modified Dirac-Pauli equation. The
generalization of the neutrino quantum equation for the case when
an electromagnetic field is present, in addition to the background
matter, is discussed below in Section 2.2 . Here we should like to
note that Eq.(\ref{D_P_matter}) is derived under the assumption
that the matter term is small. This condition is similar to the
condition of smallness of the electromagnetic term in the
Dirac-Pauli equation (\ref{D_P}) in the electrodynamics.

The detailed discussion on the evaluation of the tensor $G_{\mu
\nu}$ is given in
\cite{EgoLobStuPLB00LobStuPLB01,DvoStuJHEP02,StuPAN04}. We
consider here the case of the electron neutrino moving in the
unpolarized matter composed of the only one type of fermions of a
constant density. For a background of only electrons we  get
\begin{equation}\label{G_1}
G^{\mu \nu}= \gamma \rho^{(1)} n
\begin{pmatrix}{0}&{0}& {0}&{0} \\
{0}& {0}& {-\beta_{3}}&{\beta_{2}} \\
{0}&{\beta_{3}}& {0}&{-\beta_{1}} \\
{0}&{-\beta_{2}}& {\beta_{1}}& {0}
\end{pmatrix}, \gamma=(1-{\bm\beta}^{2})^{-1/2},
\rho^{(1)}=\frac{\tilde{G}_{F}}{2\sqrt{2}\mu},
\end{equation}
\begin{equation*}
\tilde{G}_{F}=G_F(1+4\sin^{2}\theta_{W}),
\end{equation*}
where ${\bm \beta}=(\beta_{1},\beta_{2},\beta_{3})$ is the
neutrino three-dimensional speed, $n$ denotes the number density
of the background electrons, $G_F$ is the Fermi constant, and
$\theta _{W}$ is the Weinberg angle. Note that the neutrino
magnetic moment simplifies in Eq.(\ref{D_P_matter}). From
(\ref{G_1}) and the two equations, (\ref{D_P}) and
(\ref{D_P_matter}), it is possible to see that the term ${\bf
M}=\gamma \rho^{(1)}n{\bm \beta}$ in Eq.(\ref{D_P_matter}) plays
the role of the magnetic field $\bf B$ in Eq.(\ref{D_P}).
Therefore, the Hamiltonian form of (\ref{D_P_matter}) is
\begin{equation}\label{H_G_form}
i\frac{\partial}{\partial t}\Psi({\bf r},t)=\hat H_{G}\Psi({\bf
r},t),
\end{equation}
where
\begin{equation}\label{H_G}
  \hat H_{G}=\hat {\bm{\alpha}} {\bf p} + \hat {\beta}m + \hat V_G,
\end{equation}
and
\begin{equation}\label{V_G}
\hat V_G= -\frac{\tilde{G}_{F}}{2\sqrt{2}}\frac{n}{m}
  \hat {\beta}{\bf \Sigma}{\bf p},
\end{equation}
here $\bf p$ is the neutrino momentum.

Let us now determine the energies of the two different neutrino
helicity states in matter. For the stationary states of
Eq.(\ref{H_G_form}) we get
\begin{equation}\label{stat_states}
\Psi({\bf r},t)=e^{-i( Et-{\bf p}{\bf r})}u({\bf p},E),
\end{equation}
where $u({\bf p},E)$ is independent on the spacial coordinates and
time. Upon the condition that Eq.(\ref{D_P_matter}) has a
non-trivial solution, we arrive to the energy spectrum of
different helicity states in the background matter
\cite{StuTer0410296_97}:
\begin{equation}\label{energy_1}
E=\sqrt{{\bf p}^{2}+m^2\Big(1-s\frac{\alpha p}{m}\Big)^{2}}.
\end{equation}
 where
\begin{equation}\label{alpha}
  \alpha=
  \frac{1}{2\sqrt{2}}{\tilde G}_{F}\frac{n}{m}.
\end{equation}
 It is important that the energy (\ref{energy_1}) in the
background matter depends on the state of the neutrino
longitudinal polarization (helicity), i.e. the negative-helicity
and positive-helicity neutrinos with equal momentum $\bf p$ have
different energies.

Note that in the relativistic energy limit the negative-helicity
neutrino state is dominated by the left-handed chiral state
($\nu_{-}\approx \nu_{L}$), whereas the positive-helicity state is
dominated by the right-handed chiral state ($\nu_{+}\approx
\nu_{R}$). For the relativistic neutrinos  one can derive, using
Eq.(\ref{V_G}), the probability of the neutrino spin oscillations
$\nu_{L} \leftrightarrow \nu_{R}$ with the correct form of the
matter term \cite{EgoLobStuPLB00LobStuPLB01,DvoStuJHEP02,StuPAN04}
(for the further details see Section 2.2).

The procedure, similar to one used for the derivation of the
solution of the Dirac equation in vacuum, can be adopted for the
case of the neutrino moving in matter. We apply this procedure to
Eq.(\ref{H_G_form}) and arrive to the final form of the wave
function of a neutrino moving in the background matter
\cite{StuTer0410296_97}:
\begin{equation}\label{wave_function_1}
\Psi_{{\bf p},s}({\bf r},t)=\frac{e^{-i( Et-{\bf p}{\bf
r})}}{2L^{\frac{3}{2}}}
\begin{pmatrix}{\sqrt{1+ \frac{m-s\alpha p}{E}}}
\ \sqrt{1+s\frac{p_{3}}{p}}
\\
{s \sqrt{1+ \frac{m-s\alpha p}{E}}} \ \sqrt{1-s\frac{p_{3}}{p}}\ \
e^{i\delta}
\\
{  s\sqrt{1- \frac{m-s\alpha p}{E}}} \ \sqrt{1+s\frac{p_{3}}{p}}
\\
{\sqrt{1- \frac{m-s\alpha p}{E}}} \ \ \sqrt{1-s\frac{p_{3}}{p}}\
e^{i\delta}
\end{pmatrix} ,
\end{equation}
where $L$ is the normalization length and $ \delta=\arctan
p_{y}/p_{x}$. In the limit of vanishing density of matter, when
$\alpha\rightarrow 0$, the wave function of
Eq.(\ref{wave_function_1}) transforms to the  solution of the
Dirac equation in the vacuum.

Calculations on the basis of the modified Dirac-Pauli equation
(\ref{D_P_matter}) enables us to reproduce, to the lowest order of
the expansion over the matter density, the correct energy
difference between the two neutrino helicity states in matter.
Therefore, the quantum theory of the $SL\nu$ in the lowest
approximation over the matter density can be developed using this
equation (see Section 3). However, in order to derive the correct
absolute values for the two neutrino helicity states in matter, we
investigate in Section 2.3 the neutrino quantum states on the
basis of the modified Dirac equation that we obtain from the
corresponding neutrino interaction Lagrangian.

\subsection{Modified Dirac-Pauli equation in magnetized matter}

We should like to note that it is easy to generalize the
Dirac-Pauli equation (\ref{D_P}) (or (\ref{D_P_matter})) for the
case when a neutrino is moving in the magnetized background matter
\cite{StuTer0410296_97}. For this case (i.e., when the effects of
the presence of matter and a magnetic field on neutrino have to be
accounted for) the modified Dirac-Pauli equation is
\begin{equation}\label{D_P_matter_1}
\Big\{ i\gamma^{\mu}\partial_{\mu} - m -\frac{\mu}{2}\sigma ^{\mu
\nu}(F_{\mu \nu}+G_{\mu \nu})\Big\}\Psi(x)=0,
\end{equation}
where the magnetic field $\bf B$ enters through the tensor
$F_{\mu\nu}$. If a constant magnetic field
  present in the background
 and a neutrino is moving parallel (or anti-parallel) to
the field vector $\bf {B}$, then the corresponding neutrino energy
spectrum can be obtained within the procedure discussed above in
the previous section. In particular, the neutrino energy and wave
function in the magnetized matter can be obtained from
(\ref{energy_1}) and (\ref{wave_function_1}) by the following
redefinition
\begin{equation}\label{redefin}
  \alpha \rightarrow \alpha '=\alpha
  +\frac{\mu B_{\parallel}}{p},
\end{equation}
where $B_{\parallel}= ({\bf B} {\bf p})/ p$. Thus, the neutrino
energy in this case reads \cite{StuTer0410296_97}
\begin{equation}\label{energy_2}
E=\sqrt{{\bf p}^{2}+m^2\Big(1-s\frac{\alpha p +\mu
B_{\parallel}}{m}\Big)^{2}}.
\end{equation}
For the relativistic neutrinos the expression of
Eq.(\ref{energy_2}) gives, in the linear approximation over the
matter density and the magnetic field strength, the correct value
(see \cite{EgoLobStuPLB00LobStuPLB01, StuPAN04}) for the energy
difference of the two opposite helicity states in the magnetized
matter:
\begin{equation}\label{Delta}
\Delta_{eff}= {{\tilde {G}}_F \over \sqrt{2}}n +2{\mu
B_{\parallel} \over \gamma}.
\end{equation}
that confirms our previous result of
refs.\cite{EgoLobStuPLB00LobStuPLB01, StuPAN04}.

\subsection{Modified Dirac equation for neutrino in
matter}\label{D_matter}

 The absolute value of the energy of the neutrino helicity states
in the presence of matter can be obtained on the basis of the
modified Dirac equation that we derive below directly from the
neutrino interaction Lagrangian. For definiteness, we consider
again the case of the electron neutrino propagating through moving
and polarized matter composed of only electrons (the electron
gas). The generalizations for the other flavour neutrinos and also
for more complicated matter compositions are just straightforward.

Assume that the neutrino interactions are described by the
extended standard model supplied with $SU(2)$-singlet right-handed
neutrino $\nu_{R}$. We also suppose that there is a macroscopic
amount of electrons in the scale of a neutrino de Broglie wave
length. Therefore, the interaction of a neutrino with the matter
(electrons) is coherent. In this case the averaged over the matter
electrons addition to the vacuum neutrino Lagrangian, accounting
for the charged- and neutral-current interactions, can be written
in the form
\begin{equation}\label{Lag_f}
\Delta L_{eff}=-f^\mu \Big(\bar \nu \gamma_\mu {1+\gamma^5 \over
2} \nu \Big), \ \  f^\mu={G_F \over \sqrt2}\Big((1+4\sin^2 \theta
_W) j^\mu - \lambda ^\mu \Big),
\end{equation}
where the electrons current $j^{\mu}$ and electrons polarization
$\lambda^{\mu}$ are given by
\begin{equation}
j^\mu=(n,n{\bf v}), \label{j}
\end{equation}
and
\begin{equation} \label{lambda}
\lambda^{\mu} =\Bigg(n ({\bm \zeta} {\bf v} ), n {\bm \zeta}
\sqrt{1-v^2}+ {{n {\bf v} ({\bm \zeta} {\bf v} )} \over
{1+\sqrt{1- v^2}}}\Bigg).
\end{equation}

 The Lagrangian (\ref{Lag_f})
accounts for the possible effect of the matter motion and
polarization. Here $n$, ${\bf v}$, and ${\bm \zeta} \ (0\leqslant
|{\bm \zeta} |^2 \leqslant 1)$ denote, respectively, the number
density of the background electrons, the speed of the reference
frame in which the mean momentum of the electrons is zero, and the
mean value of the polarization vector of the background electrons
in the above mentioned reference frame. The detailed discussion on
the determination of the electrons polarization can be found in
\cite{EgoLobStuPLB00LobStuPLB01,DvoStuJHEP02,StuPAN04}.

From the standard model Lagrangian with the extra term $\Delta
L_{eff}$ being added, we derive the following modified Dirac
equation \cite{StuTer0410296_97} for a neutrino moving in the
background matter,
\begin{equation}\label{new}
\Big\{ i\gamma_{\mu}\partial^{\mu}-\frac{1}{2}\gamma_{\mu}
(1+\gamma_{5})f^{\mu}-m \Big\}\Psi(x)=0.
  \end{equation}
This is the most general equation of motion of a neutrino in which
the effective potential $V_{\mu}=\frac{1}{2}(1+\gamma_{5})f_{\mu}$
accounts for both the charged and neutral-current interactions
with the background matter and also for the possible effects of
the matter motion and polarization. It should be noted here that
the modified effective Dirac equations for a neutrino with various
types of interactions with the  background environment  were used
previously in \cite{ManPRD88, PanPLB91-PRD92, ChaZiaPRD88,
NotRafNPB88, NiePRD89, HaxZhaPRD91,WeiKiePRD97} for the study of
the neutrino dispersion relations and derivation of the neutrino
oscillation probabilities in matter. If we neglect the
contribution of the neutral-current interaction and possible
effects of motion and polarization of the matter then from
(\ref{new}) we can get corresponding equations for the left-handed
and right-handed chiral components of the neutrino field derived
in \cite{PanPLB91-PRD92}. The similar equation for a neutrino in
the background of non-moving and unpolarized neutrons was also
used in \cite{KachPLB98, KusPosPLB02}.

Upon the condition that the equation (\ref{new}) has a non-trivial
solution, we arrive to the energy spectrum of a neutrino moving in
unpolarized background matter at rest:
\begin{equation}\label{energy_new}
  E_{\varepsilon}=\varepsilon{\sqrt{{\bf p}^{2}\Big(1-s\alpha \frac{m}{p}\Big)^{2}
  +m^2} +\alpha m}.
\end{equation}
The quantity $\varepsilon=\pm 1$ splits the solutions into the two
branches that in the limit of the vanishing matter density,
$\alpha\rightarrow 0$, reproduce the positive and
negative-frequency solutions, respectively. Note that again the
neutrino energy in the background matter depends on the state of
the neutrino longitudinal polarization, i.e. in the relativistic
case the left-handed and right-handed neutrinos with equal momenta
have different energies.

 Although the obtained neutrino energy
spectrum (\ref{energy_new}) does not reproduce the one of
Eq.(\ref{energy_1}), an equal result for the energy difference
$\Delta E=E(s=-1)-E(s=+1)$ of the two neutrino helicity states can
be obtained from both of the spectra in the low matter density
limit $\alpha\frac {pm}{E_{0}^{2}} \ll 1$:
\begin{equation}\label{delta_Energy}
  \Delta E\approx 2m\alpha \frac{p}{E_{0}},
\end{equation}
where we use the notation $E_0=\sqrt{p^2 +m^2}$. It should be also
noted that for the relativistic neutrinos the energy spectrum for
the neutrino helicity states of Eq.(\ref{energy_new}) in the low
density limit reproduces the correct energy values for the
neutrino left-handed and right-handed chiral states:
\begin{equation}\label{E_L}
  E_{\nu_L} \approx E(s=-1)\approx E_0 +{{\tilde {G}}_F \over \sqrt{2}}n,
\end{equation}
and
\begin{equation}\label{E_R}
  E_{\nu_R} \approx E(s=-1)\approx E_0,
\end{equation}
as it should be for the active left-handed and sterile
right-handed neutrino in matter.

For the wave function of a neutrino in the background matter given
by Eq.(\ref{new}) we get:
\begin{equation}\label{wave_function}
\Psi_{\varepsilon, {\bf p},s}({\bf r},t)=\frac{e^{-i(
E_{\varepsilon}t-{\bf p}{\bf r})}}{2L^{\frac{3}{2}}}
\begin{pmatrix}{\sqrt{1+ \frac{m}{ E_{\varepsilon}-\alpha m}}}
\ \sqrt{1+s\frac{p_{3}}{p}}
\\
{s \sqrt{1+ \frac{m}{ E_{\varepsilon}-\alpha m}}} \
\sqrt{1-s\frac{p_{3}}{p}}\ \ e^{i\delta}
\\
{  s\varepsilon\sqrt{1- \frac{m}{ E_{\varepsilon}-\alpha m}}} \
\sqrt{1+s\frac{p_{3}}{p}}
\\
{\varepsilon\sqrt{1- \frac{m}{ E_{\varepsilon}-\alpha m}}} \ \
\sqrt{1-s\frac{p_{3}}{p}}\ e^{i\delta}
\end{pmatrix} ,
\end{equation}
Obviously, in the limit of vanishing density of matter, when
$\alpha\rightarrow 0$, the wave function (\ref{wave_function})
transforms to  the solution of the Dirac equation for a neutrino
in the vacuum.

\section{Spin light of neutrino in matter and magnetic field}

The proposed  quantum equations
(\ref{D_P_matter}),(\ref{D_P_matter_1}) and (\ref{new}) for a
neutrino moving in the background matter establish a basis for a
new method in the investigation  of different processes with
participation of neutrinos in the presence of matter and external
electromagnetic fields. As an example, we should like to study of
the ($SL\nu$) in the magnetized matter and  develop the {\it
quantum} theory of this effect.

Within the quantum approach, the corresponding Feynman diagram of
the $SL\nu$ in matter is the standard one-photon emission diagram
with the initial and final neutrino states described by the "broad
lines" that account for the neutrino interaction with matter and
the external electromagnetic field. From the usual neutrino
magnetic moment interaction, it follows that the amplitude of the
transition from the neutrino initial state $\psi_{i}$ to the final
state $\psi_{f}$, accompanied by the emission of a photon with a
momentum $k^{\mu}=(\omega,{\bf k})$ and a polarization ${\bf
e}^{*}$, can be written in the form
\begin{equation}\label{amplitude}
  S_{f i}=-\mu {\sqrt {\frac {2\pi}{\omega L^{3}}}}
  2\pi\delta(E_{f}-E_{i}+\omega)
  \int d^{3} x {\bar \psi}_{f}({\bf r})({\hat {\bf \Gamma}}{\bf e}^{*})
  e^{i{\bf k}{\bf r}}
   \psi_{i}({\bf r}),
\end{equation}
where
\begin{equation}\label{Gamma}
  \hat {\bm \Gamma}=i\omega\big\{\big[{\bm \Sigma} \times
  {\bm \varkappa}\big]+i\gamma^{5}{\bm \Sigma}\big\}.
\end{equation}
Here ${\bm \varkappa}=\frac{\bf k}{\omega}$ is the unit vector
pointing in the direction of the emitted photon propagation. The
delta-function stands for the energy conservation. Performing the
integrations over the spatial co-ordinates, we can recover the
delta-functions for the three components of the momentum. In the
lowest order of the expansion over the density of the background
matter, the properties of the $SL\nu$ (in particular, the rate and
radiation power), obtained on the basis of Eqs. (\ref{D_P_matter})
and (\ref{new}), are the same. This is because, as it has been
already mentioned, the difference of the energies of the two
neutrino helicity states, calculated with use of Eqs.
(\ref{D_P_matter}) and (\ref{new}), are equal. The additional
effect from a magnetic field (if it is also present in the
background environment) can be accounted for if one describes the
neutrino on the basis the modified Dirac-Pauli equation of
Eq.(\ref{D_P_matter_1}). Note that the $SL\nu$ in the presence of
a magnetic field can have interesting applications for magnetized
astrophysical media.

Let us suppose that the weak interaction of the neutrino with the
electrons of the background is indeed weak. Thus, in wide ranges
of densities of matter and strengths of the magnetic field that
are appropriate for the astrophysical applications,  we can expand
the energy (\ref{energy_2}) over $\alpha'\frac {pm}{E_{0}^{2}} \ll
1$ and in the liner approximation get for the emitted photon
energy
\begin{equation}\label{omega_1}
    \omega=(s_{f}-s_{i})\alpha' m  \frac{\beta}{1-\beta \cos
    \theta},
\end{equation}
where $\theta$ is the angle between ${\bm \varkappa}$ and the
direction of the neutrino speed ${\bm \beta}$.

From the above consideration it follows that the only possibility
for the $SL\nu$ to appear is provided in the case when the
neutrino initial and final states are characterized by $s_{i}=-1$
and $s_{f}=+1$, respectively. Thus we conclude, on the basis of
the quantum treatment of the $SL\nu$ in the magnetized matter,
that in this process the relativistic left-handed neutrino is
converted to the right-handed neutrino  and the emitted photon
energy is given by
\begin{equation}\label{omega_2}
    \omega=
    \frac {1}{1-\beta \cos
    \theta}\omega_0,
\end{equation}
where we use the notation
\begin{equation}\label{omega_0}
\omega_0= \frac {{\tilde G}_{F}} {\sqrt{2}}n\beta+
    2\frac {\mu B_{\parallel}}{\gamma}.
\end{equation}
Note that the photon energy depends on the angle $\theta$ and also
on the value of the neutrino speed $\beta$. In the case of
$\beta\approx 1$ and $\theta\rightarrow 0$ we confirm the
estimation for  the emitted photon energy in the background matter
obtained in \cite{LobStuPLB03}. If the effect of the background
matter is subdominant and the main contribution to the $SL\nu$ is
given by the magnetic field term,  then from Eq.(\ref{omega_2}) we
obtain the corresponding result of ref.\cite{BorZhuTer88} where
the $SL\nu$ in the presence of a magnetic field was considered.

For the spin light transition rate in the lowest order
approximation over the parameter $\alpha'\frac {pm}{E_{0}^{2}}$ we
get \cite{StuTer0410296_97}
\begin{equation}\label{rate}
  \Gamma_{SL}=
  \mu^{2}\omega_0^3
    \int
  \limits_{}^{}\frac{S\sin \theta}{(1-\beta \cos \theta)^{4}}
   d\theta,
\end{equation}
where
\begin{equation}\label{S}
S=(\cos \theta - \beta)^{2}+
  (1-\beta \cos \theta)^{2}.
\end{equation}
The corresponding expression for the radiation power is
\begin{equation}\label{power}
  I_{SL}=\mu^{2}\omega_0^4
  \int\limits_{}^{}\frac{S\sin \theta}{(1-\beta \cos \theta)^{5}}
   d\theta.
\end{equation}
Performing the integrations in Eq.(\ref{rate}) over the angle
$\theta$, we obtain for the rate
\begin{equation}\label{rate_1}
  \Gamma_{SL}=\frac{8}{3}\mu^{2}
  \omega_0^{3}\gamma^{2}.
\end{equation}
 For the total radiation power from Eq.(\ref{power}) we get,
\begin{equation}\label{power_1}
  I_{SL}=\frac{8}{3}\mu^{2}\omega_0 ^{4}\gamma^{4}.
\end{equation}

From the obtained result of Eq.(\ref{rate_1}) by switching off the
contribution from the magnetic field one can get the $SL\nu$ rate
in matter (see also \cite{LobStuPLB04}). This result exceeds the
value of the neutrino spin light rate derived in
\cite{LobStuPLB03} by a factor of two because here the neutrinos
in the initial state  are totally left-handed polarized, whereas
in \cite{LobStuPLB03} the case of initially unpolarized neutrinos
(i.e., an equal mixture of the left- and right-handed neutrinos)
is considered.

\section{Summary and conclusion}
We have developed the quantum approach to description of a
neutrino moving in the background matter on the basis of the
generalized Dirac-Pauli and modified Dirac equations. In the low
matter density limit the two equations give equal values for the
energy difference of the opposite neutrino helicity states in
matter. The use of the generalized Dirac-Pauli equation also
enables us to account for the effect of the longitudinal magnetic
field. In the derivation of the Dirac-Pauli equation (see
(\ref{D_P_matter}) and (\ref{D_P_matter_1}) ) it has been supposed
that the matter parameter is small,
\begin{equation}\label{restriction}
\alpha\frac {pm}{E_{0}^{2}}= \frac{1}{2\sqrt{2}}{\tilde
G}_{F}\frac{np}{E_{0}^{2}}\ll 1.
\end{equation}
However, due to the fact that even for the extremely dense matter
with $n=10^{37} \ \ cm^{-3}$ one gets $\frac{1}{2\sqrt{2}}{\tilde
G}_{F}n\sim 1 \ eV$, the restriction (\ref{restriction}) does not
forbid to use the generalized Dirac-Pauli equation even for a very
dense medium of neutron stars if the neutrino has the relativistic
energy. It should be noted that in the derivation of the modified
Dirac equation (\ref{new}) no any restrictions of this kind has
been made.

On the basis of these two equations the quantum treatment of a
neutrino moving in the presence of the background matter has been
realized with the effect of the longitudinal magnetic field being
incorporated. Within the developed quantum approach, the emission
rate and power of the $SL\nu$ in magnetized matter has been
calculated accounting for the emitted photons polarization. The
existence of the neutrino-self polarization effect in the process
of the spin light radiation in the background matter and magnetic
field has been shown. The photon energy, in the case when the both
effects of the background matter and longitudinal magnetic field
are important, has been derived for the first time. The photon
energy depends on the density of matter, the value of the neutrino
magnetic moment, the strength of the magnetic field and also on
the direction of the neutrino propagation in respect to the
magnetic field $\bm B$ orientation. The $SL\nu$ radiation and the
corresponding neutrino self-polarization effect, due to the
significant dependence on the matter density and the magnetic
field strength, are expected to be important in different
astrophysical dense media and in the early Universe.

In conclusion, let us consider the case when the background
magnetic field is strong enough so that the following condition is
valid
\begin{equation}\label{str_B}
  B>\frac{p_{F}^{2}}{2e},
\end{equation}
where $p_{F}=\sqrt{\mu ^2 - m^{2}_{e}}$, $\mu$ and $m_{e}$ are,
respectively, the Fermi momentum, chemical potential, and mass of
electrons. Then all of the electrons of the background occupy the
lowest Landau level (see, for instance, \cite{NunSemSmiValNPB97}),
therefore the matter is completely polarized in the direction
opposite to the unit vector ${{\bm B}}/{B}$. From the general
expression for the tensor $G_{\mu \nu}$ (see the second paper of
\cite{EgoLobStuPLB00LobStuPLB01}) we get
\begin{equation}\label{G_1}
G^{\mu \nu}= \gamma  n \left\{
\rho^{(1)}\begin{pmatrix}{0}&{0}& {0}&{0} \\
{0}& {0}& {-\beta_{3}}&{\beta_{2}} \\
{0}&{\beta_{3}}& {0}&{-\beta_{1}} \\
{0}&{-\beta_{2}}& {\beta_{1}}& {0}
\end{pmatrix}
+
\rho^{(2)}\begin{pmatrix}{0}&{-\beta_{2}}& {\beta_{1}}&{0} \\
{\beta_{2}}& {0}& {-\beta_{0}}&{0} \\
{-\beta_{1}}&{\beta_{0}}& {0}&{0} \\
{0}&{0}& {0}& {0}
\end{pmatrix}\right\},
\end{equation}
\begin{equation*}
\rho^{(2)}=-\frac{{G}_{F}}{2\sqrt{2}\mu}.
\end{equation*}
On the basis of the modified Dirac-Pauli equation
(\ref{D_P_matter_1}) with the tensor $G_{\mu \nu}$ given by
(\ref{G_1}) it is possible to consider the $SL\nu$ in the case
when the neutrino is moving in the completely polarized matter
parallel (or anti-parallel) to the magnetic field vector $\bm B$.
The neutrino energy and wave function in such a case can be
obtained from (\ref{energy_1}) and (\ref{wave_function_1}) by the
following redefinition
\begin{equation}\label{redefin}
  \alpha \rightarrow {\tilde \alpha} =\alpha
  \left[1- \frac{sign \left(\frac{B_{\parallel}}{B}\right)}
  {1+\sin^{2}
  4\theta_{W}}\right]
  +\frac{\mu B_{\parallel}}{p}.
\end{equation}
The second term in brackets in Eq.(\ref{redefin}) accounts for the
effect of the matter polarization. It follows, that the effect of
the matter polarization can reasonably change the total matter
contribution to the neutrino energy (\ref{energy_2}). The emitted
$SL\nu$ photon energy is determined in this case by
(\ref{omega_2}) with
\begin{equation}\label{omega_0_1}
\omega_0= \frac {{ G}_{F}} {\sqrt{2}}n \beta\left[
\left(1+\sin^{2}
  4\theta_{W}\right)-{sign
\left(\frac{ B_{\parallel}}{B}\right)}\right]+
    2\frac {\mu B_{\parallel}}{\gamma}.
\end{equation}
Consequently, the effect of the matter polarization significantly
influence the rate and radiation power of the $SL\nu$.

\end{document}